\documentclass[letterpaper,10pt]{article} 
\usepackage{opticameet3} 
\newcommand\authormark[1]{\textsuperscript{#1}}
\abovedisplayshortskip 
\belowdisplayshortskip
\usepackage{amsmath,amssymb}
\usepackage{graphicx}
\usepackage{subfigure}
\usepackage{booktabs}
\usepackage{caption}
\usepackage{svg}
\usepackage[colorlinks=true,bookmarks=false,citecolor=blue,urlcolor=blue]{hyperref} 
\usepackage{float}
\begin{document}
\vspace{-15pt}
\title{A Novel Machine Learning-based Equalizer for a Downstream 100G PAM-4 PON}

\copyrightyear{2024}

\author{Chen Shao\authormark{1*}, Elias Giacoumidis\authormark{2*}, Shi Li\authormark{2}, Jialei Li\authormark{3}, Michael Färber\authormark{1}, Tobias Käfer\authormark{1}, and André Richter\authormark{2}}

\address{
\authormark{1} Karlsruhe Institute of Technology (KIT), Kaiserstr.\ 12, 76131 Karlsruhe, Germany\\
\authormark{2} VPIphotonics GmbH, Hallerstr.\ 6, 10587 Berlin, Germany\\
\authormark{3} RWTH Aachen University, Templergraben 57, 52062 Aachen\\
}
\email{\authormark{*}chen.shao2@kit.edu, \authormark{*}elias.giacoumidis@VPIphotonics.com} 

\vspace{-15pt}
\begin{abstract}
A frequency-calibrated SCINet (FC-SCINet) equalizer is proposed for downstream 100G PON with 28.7 dB path loss. At 5 km, FC-SCINet improves the BER by $~$88.87\% compared to FFE and a 3-layer DNN with 10.57\% lower complexity.
\end{abstract}

\vspace{0pt} 
\section{Introduction}
\label{sec: intro}
\vspace{-5pt} 

Data traffic in access networks experiences growth, requiring passive optical networks (PONs) data rates beyond 10 Gb/s \cite{Kaneda2020FPGA}. Although a 50G intensity-modulated and direct-detected (IMDD) PON standard was recently agreed upon, research efforts are already ongoing for 100 Gb/s \cite{Borkowski2020PON}. For such PON, chromatic dispersion (CD) produces nonlinear distortion in the received signal that is best handled by digital signal processing such as feed-forward/decision-feedback equalizers (FFE/DFE). A practical 100G PON implementation is based on 25G-class electro-absorption modulation (EAM) with semiconductor optical amplification and pulse amplitude modulation (PAM-4) \cite{Borkowski2020PON}. Using FFE/DFE with implementable filter lengths, i.e., 23 feed-forward and 2 decision-feedback taps, an optical path loss (OPL) of $>31.5$ dB is supported due to powerful forward-error-correction (FEC).

Machine learning-based equalization methods, such as deep neural networks (DNN) \cite{Yi2019ML100G,Kaneda2020FPGA}, have revealed similar performance as FFE and Volterra equalizers under linear distortions but stronger performance in the nonlinear regime \cite{Yi2019ML100G}. At a launch optical power (LOP) of 18 dBm using a booster amplifier in the optical line terminal (OLT), a 30 dB loss budget was feasible in DNN-assisted 100G IMDD PONs using 20G-class electro-optical devices. In \cite{Yi2019ML100G}, a Mach-Zehnder modulator was adopted. However, practical PONs use an EAM that introduces chirp, causing temporal pulse spreading. Besides CD, potential residual time jitter can be detrimental in PONs induced from linewidth broadening and frequency fluctuations in the laser source or mechanical vibrations introducing physical disturbances affecting the stability of optical components. 

This paper introduces FC-SCINet, a novel equalizer that integrates frequency-calibration (FC) into a sampling convolutional and interaction network (SCINet) for a downstream 100G PAM-4 PON with 28.7 dB OPL. By combining FC with SCINet, FC-SCINet not only enables effective time-series modeling with intricate temporal dynamics \cite{Liu2021SCINetTS} but also supports fine-tuning of signal spectral attributes. FC-SCINet demonstrates a significant enhancement in CD tolerance, surpassing the performance of a 9/21-tap FFE and a 3-layer DNN by achieving a minimum of 87.5\% bit-error-rate (BER) improvement at 11 km transmission. FC-SCINet exhibits robustness against EAM chirp, residual time jitter, and Kerr nonlinearity, delivering 88.87\% BER improvement (at a FEC limit of $10^{-2}$) compared to FFE and DNN at 5 km, while maintaining a 10.577\% lower complexity.

\vspace{-5pt} 
\section{Evaluation Setup} 
\label{sec:simulation setup}
\vspace{-5pt} 
We used VPItransmissionMaker to simulate a downstream 100G PON at 1550 nm. In the OLT transmitter, an EAM was employed at 55 GHz bandwidth with realistic transmittance and chirp characteristics identical to \cite{Zhang2018Performance}, driven by an electrical PAM-4 signal generator that uses rectangular pulse shaping. For the transmitter laser, we set: a linewidth of 1 MHz, side-mode separation of 200 GHz, side-mode suppression ratio of 100 dB, and a relative intensity noise of -130 dB/Hz. The LOP was fixed at 10 dBm to account for a total OPL of 28.7 dB. For the optical network unit (ONU) receiver, an avalanche photo-diode at 55 GHz bandwidth was simulated with thermal noise of $10^{-12}$ A/sqrt(Hz) including shot noise, responsivity of 0.9 A/W, multiplication factor of 8 and ionization coefficient of 0.4; followed by a transimpedance amplifier with a transimpedance of 100 ohms and an electrical post-amplifier with 10 dB gain. A current noise spectral density of $20^{-12}$ A/sqrt(Hz) was included for both the transmitter driver and post-amplifier. Transceiver electrical filters were modeled with a 4th-order Bessel type. To prevent bit pattern recognition and thus performance overestimation when comparing FC-SCINet and a low-complex DNN, we employed a random-unrepeated number sequence (RNS). For the DNN, we explored hidden layers ranging from 2 to 16, each with varying numbers of neurons per layer, spanning from 4 to 128. Sigmoid and ReLU activation functions were compared. The sliding time window approach was assessed using different pre- and post-window sizes (4 to 64). Other DNN parameters included a batch size of 32, learning rate of $10^{-5}$, norm-2 regularization, and mean-square error (MSE) as a loss function. The training data set for both FC-SCINet and DNN was 2$^{20}$  × 30 RNS iterations for each fiber distance, synchronized and downsampled at 1 sample-per-symbol (SpS). 15\% of the data was used for testing and 10\% for validation. For comparison, a traditional, low-complex, decision-directed FFE with 9 and 21 finite impulse response (FIR) taps was simulated, using the least-mean square algorithm to adapt the filter coefficients. For the transmission link, we consider two scenarios: 1) including only chromatic dispersion (\textbf{CD}); and 2) a more realistic case with all impairments included, e.g., CD, Kerr-induced nonlinearity, transmitter chirp and time jitter, named \textbf{Realistic}. A standard single-mode fiber was considered in both scenarios with 0.2 dB/km loss, 16 ps/nm/km CD coefficient, and 8 ps/nm/km CD-slope. For Case 2), a Kerr nonlinear coefficient of $2.6^{-20}$ m$^2$/W was considered. Timing jitter was emulated using Gaussian distribution for each generated symbol with a root-mean-square of 0.1 and a 0.5 standard deviation. Jitter emulation caused level transitions to shift in time, and was not applied to the leftmost and rightmost transitions. 

\begin{figure*}[t!]
  \centering
  \captionsetup{font=footnotesize} 
  \subfigure[]
  {
    \includegraphics[width=0.77\textwidth]{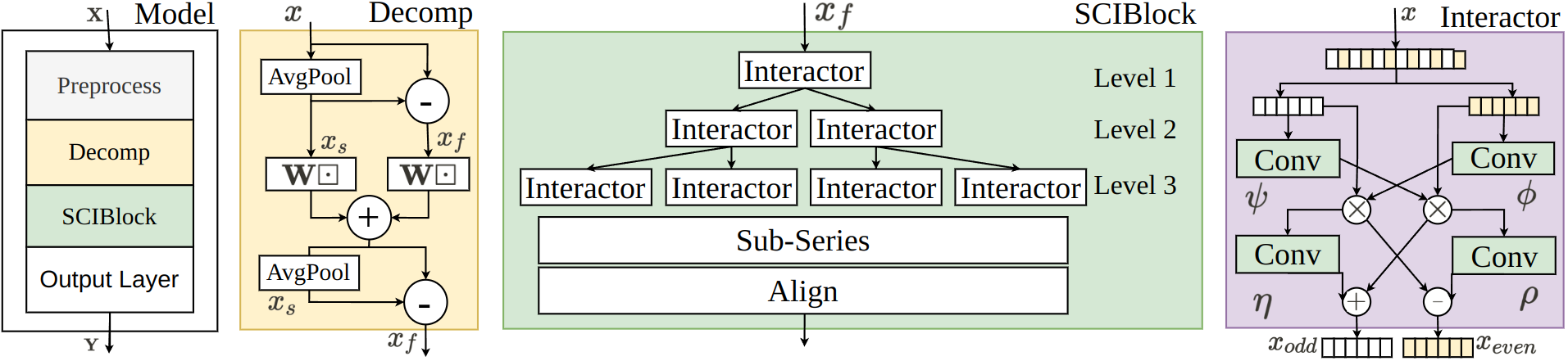}
    \label{subfig:1a}
    }
  \subfigure[]
  {
    \includegraphics[width=0.19\textwidth]{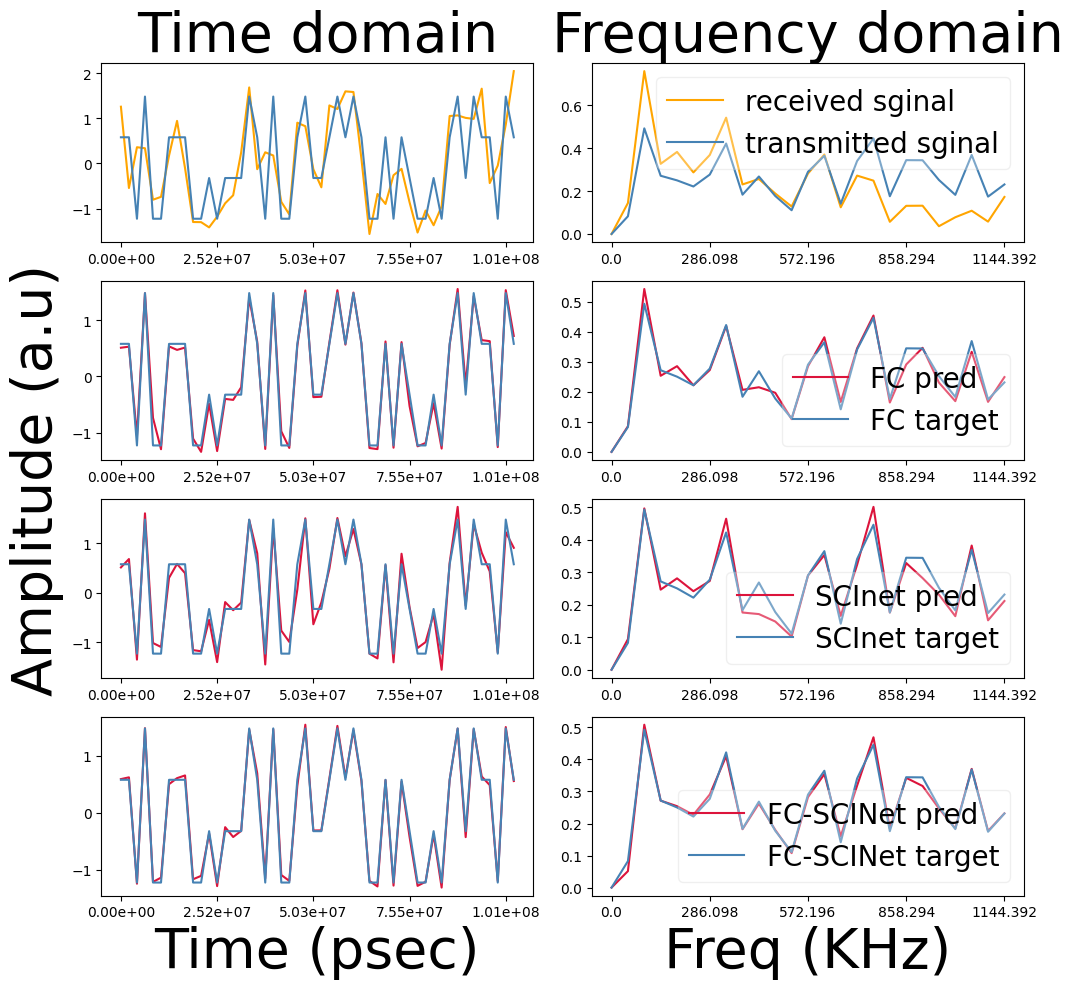}
    \label{subfig:1b}
    }
      \vspace{-10pt} 
  \caption{(a) Block diagram illustrating the innovative FC-SCINet equalizer. (b) Time/Frequency domain representation of 50 transmitted (target)/received consecutive samples, showing separately the impact of FC and SCInet.}  
  \vspace{-5pt}
  \label{fig:Fig1}
\end{figure*}


\vspace{-5pt} 
\section{FC-SCINet Model}
\label{sec:model}
\vspace{-5pt} 

The objective of FC-SCINet is to achieve an optimal mapping by capturing the pattern of distortion caused by channel/transceiver effects in the received signal X. This enables the prediction of the transmitted signal Y. FC-SCINet comprises 3 key components: Preprocess, Decomp, and SCIBlock, as illustrated in Fig. \ref{subfig:1a}.

\noindent \textbf{Preprocess}: The Preprocess module in Fig. \ref{subfig:1a} consists of 3 sub-modules: normalization, windowing (sliding window), and frequency-calibration (FC). The dataset is firstly normalized to have 0 mean and a standard deviation of 1. Afterward, the normalized input X is batched into a tensor with a fixed window length $p$. Then, we perform FC to align spectral components with the desired reference. We decompose the input sequence into smooth $\mathbf{x_{s}}$ and fluctuating components $\mathbf{x_{f}}$ with a moving average filter AvgPool, and then take $\mathbf{x_{pre}}$ as input for further processing according to Eq.\eqref{eqn:equ1}.
\vspace{-10pt} 
\begin{align}
\label{eqn:equ1}
  \mathbf{x_s} = \text{AvgPool}(\mathbf{x})  \quad 
  \mathbf{x_f} =  \mathbf{x} -  \mathbf{x_s}  \quad 
  \mathbf{x}_{pre}= \mathbf{x} + \mathbf{x_f}
\end{align}

\noindent \textbf{Decomp}: The Decomp layer is a variant of the FC module, separating $\mathbf{x_{pre}}$ into smooth and fluctuating components and then processing them through linear layers with the weight matrices $\mathbf{W_{f}}$ and $\mathbf{W_{s}}$ respectively (Eq.\eqref{eqn:equ2}). Afterward $\mathbf{\hat{x}}$ is decomposed again into $\mathbf{x_{s}}$ and  $\mathbf{x_{f}}$ for subsequent processing. 
\vspace{-5pt}
\begin{align}
\label{eqn:equ2}
    \mathbf{\hat{x}} = \mathbf{W_s^T}\mathbf{x_s} + \mathbf{W_f^T}\mathbf{x_f}
    \vspace{-2pt}
\end{align}

\noindent \textbf{SCIBlock}: The fluctuated signal $\mathbf{x_{f}}$ is processed within SCIBlock, while the smoothing-component $\mathbf{x_{s}}$ is subsequently added to the output of SCIBlock. SCIBlock features a binary tree structure, where each node corresponds to an Interactor as shown in Fig.\ref{subfig:1a} \cite{Liu2021SCINetTS}. Within each Interactor, the input signal undergoes a splitting process where the even-positioned ($\mathbf{x_{even}}$) and odd-positioned bits ($\mathbf{x_{odd}}$) (Eq. \eqref{eqn:equ3}, \eqref{eqn:equ4}) are extracted. Here, $\odot$ represents an element-wise product, and $\mathbf{\psi}$, $\mathbf{\phi}$, $\mathbf{\eta}$, and $\mathbf{\rho}$ are independent 1D convolutional layers. The intermediate outputs are denoted as $\mathbf{x_{even}^{s}}$, $\mathbf{x_{odd}^{s}}$, $\mathbf{x_{even}'}$, and $\mathbf{x_{odd}'}$.
\vspace{-5pt}
\begin{align}
\label{eqn:equ3}
  \mathbf{x_{even}^{s}} = \mathbf{x_{even}} \odot \exp(\psi(\mathbf{x_{odd}})), \quad  \mathbf{x_{odd}^{s}} = \mathbf{x_{odd}} \odot \exp(\phi(\mathbf{x_{even}})) \\
\label{eqn:equ4}
  \mathbf{x_{odd}'} = \mathbf{x^s_{even}} + \exp(\eta(\mathbf{x^{s}_{odd}})) \quad  
  \mathbf{x_{even}'} = \mathbf{x^{s}_{odd}} - \exp(\rho(\mathbf{x^{s}_{even}})) 
\end{align}
This process recurs at each level. In the final layers, the resulting sub-series $\mathbf{\hat{x_{f}}}$ are aligned based on their initial coordinates within $\mathbf{x_{f}}$.

\noindent \textbf{Output Layer}: The output $\mathbf{\hat{x_{f}}}$ of SCIBlock is added to the smoothing-component $\mathbf{x_{s}}$, which is followed by a final convolutional layer to generate the final prediction $\mathbf{\hat{y}}$. 

\noindent \textbf{Training}: In a manner akin to DNN, an MSE function, norm-2 regularization, and a $10^{-5}$ learning rate were used. 


\vspace{-5pt} 
\section{Results and Discussion}
\label{sec:result}
\vspace{-5pt} 

\begin{figure}[hptb]
  \centering
  \captionsetup{font=footnotesize} 
  \subfigure[]
  {
    \includegraphics[width=0.23\textwidth]{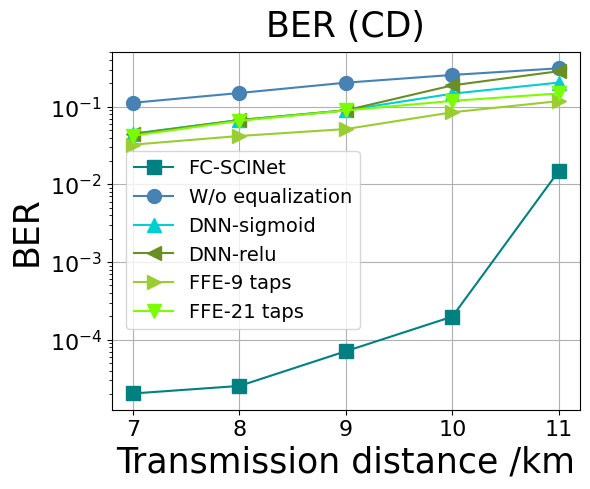}
    \label{subfig:2a}
    }
  \subfigure[]
  {
    \includegraphics[width=0.23\textwidth]{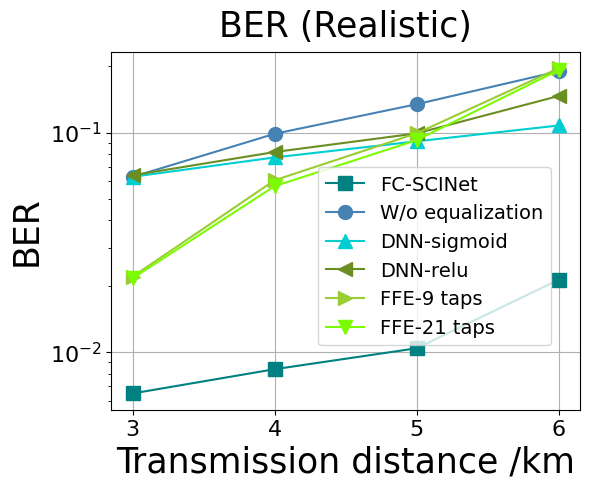}
    \label{subfig:2b}
    }
  \subfigure[]
  {
    \includegraphics[width=0.23\textwidth]{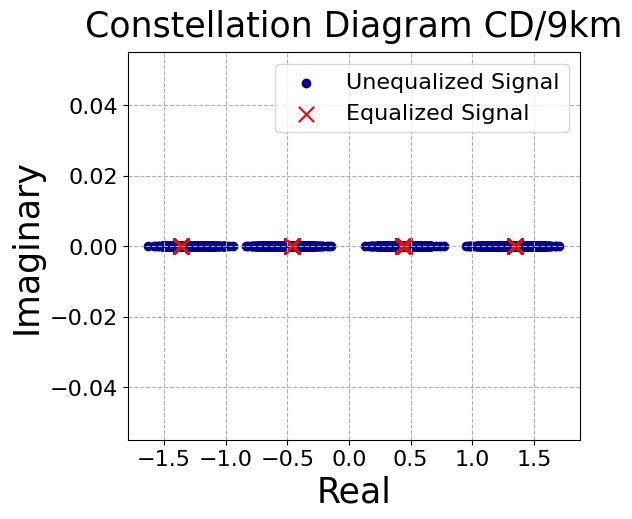}
    \label{subfig:2c}
    }
  \subfigure[]
  {
    \includegraphics[width=0.23\textwidth]{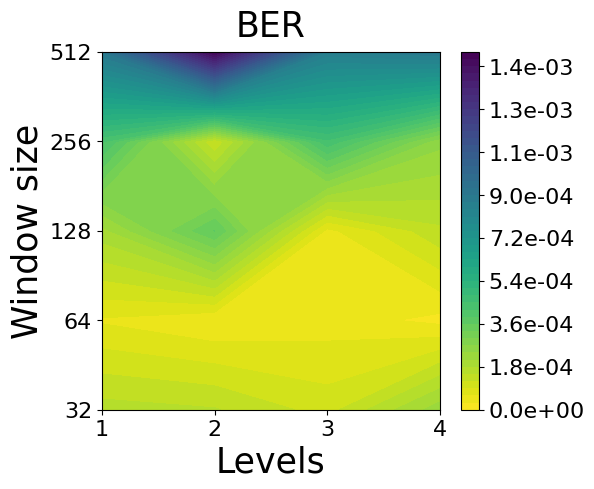}
    \label{subfig:2d}
    }
      \vspace{-10pt} 
  \caption{(a),(b) BER vs. distance for FC-SCInet, DNN, FFE and w/o equalization for Cases 1-2 (CD-Realistic). (c) Received constellation diagram for Case 1 (CD) at 9 km. (d) BER color map for FC-SCInet's window size and levels of Interactors for Case 1 (CD) at 9 km.}  
  \vspace{-10pt}
  \label{fig:Fig2}
\end{figure}

\color{black} FC-SCINet performance in terms of direct-counting-BER (Monte-Carlo) was compared with that of DNN/FFE at 1 SpS and 2$^{20}$ number of symbols for various distances with fixed 28.7 dB OPL. The DNN optimization for each distance yielded a pre- and post-window size of 16, along with 3 hidden layers comprising 60, 64, and 18 neurons, respectively. A higher number of layers led to overestimation when using ReLU or sigmoid activation functions, while the impact of neurons' number was almost negligible. For Case 1) \textbf{CD} scenario shown in Fig.\ref{subfig:2a}, FC-SCINet significantly improves the CD tolerance compared to FFE/DNN, estimating 87.5\% BER enhancement at 11 km compared to a 9-tap FFE and a 3-layer DNN. Fig. \ref{subfig:1b} demonstrates the effectiveness of FC-SCINet at 9 km, showcasing a random representative set of 50 consecutive time-domain samples and their corresponding frequency spectrum. The combined capabilities of FC and SCINet synergistically result in spectral shaping (peak-to-peak amplitude compression) and temporal pulse narrowing, leading to exceptional prediction accuracy with the transmitted sequence. Fig. \ref{subfig:2d} presents a BER color map at 9 km, highlighting the two crucial hyperparameters for optimization: the levels of Interactors and the window size. At this distance, the optimal BER is attained when the window size is set to 64 and the level is 3. Additionally, Fig. \ref{subfig:2c} displays the corresponding received PAM-4 constellation diagram, indicating that errors primarily occur in the high-amplitude region where FC-SCINet exerts its notable compensating influence. It is worth noting that DNN demonstrates inferior performance compared to FFE because the system is under a linear regime with a low signal-to-noise ratio \cite{Yi2019ML100G}, and DNN is typically beneficial at the nonlinear regime \cite{Yi2019ML100G}. For Case 2) \textbf{Realistic} scenario, shown in Fig.\ref{subfig:2b}, FC-SCINet shows robust performance in the presence of additional effects such as EAM chirp, residual time jitter, and Kerr nonlinearity. For a $10^{-2}$ FEC-limit reaching 5 km, 88.87\% BER improvement is feasible compared to FFE/DNN. 
\vspace{-12pt}

\vspace{-5pt}
The complexity of FC-SCINet was compared to the optimum 3-layer DNN [60, 64, 18 neurons] using the metric of Real Multiplications per Symbol (RMpS)\cite{Freire2021PVCSNNE}. The trade-off between BER and complexity was characterized by the \textit{PRB:} Product of RMpS with median-BER (mBER). An analytical expression of RMpS, PRB and their instantiation \footnote{$level$: Interactor level, $n_h$: number of features in a convolution neuron, $n_s$: window size, input length, $n_{i}$: number of neurons for $i$ DNN's layers, $n_{e}$: number of experiments, $n_{c}$: number of channels of the input data. $L:$ interactor level number}  are provided in Table \ref{tab:my_label}. From the RMpS values, it is evident that FC-SCINet can offer a 10.577\% complexity reduction compared to the adopted DNN.



\begin{table}[h]
    \centering
    \captionsetup{font=footnotesize}
    \footnotesize
    \setlength{\tabcolsep}{5pt} 
    \renewcommand{\arraystretch}{1.2} 
    \caption{Comparison of RMpS, mBER and PRB}
    \vspace{-5pt}
    \begin{tabular}{lcccc}
    \toprule
    \vspace{-2pt}
     Model & RMpS(Expression) \color{black}\color{black} & RMpS(Value) \color{black}\color{black} & mBER(CD/Real) & PRB(CD/Real) \\
    \midrule
    \vspace{-2pt}
     DNN & $ n_e(n_s \cdot n_c \cdot n_1 + n_1 \cdot n_2 + n_2 \cdot n_3 + n_3 \cdot n_o)$ & 209,700 & 0.089912/0.087272 & 18854.55/18300.94 \\
     FC-SCINet & $ n_s \cdot [(n_s \cdot n_e + n_s) + \sum_{level=1}^{L} \frac{1}{2} [1800 \cdot n_{h} +(n_{s}/2^{level})] + 30 ] $ & 187,520 & 0.000071/0.009414 & 13.31/1765.31 \\
    \bottomrule
    \end{tabular}
    \label{tab:my_label}
    \vspace{-10pt}
\end{table}

\vspace{-5pt} 
\section{Conclusion}
\label{sec:conclusion}
\vspace{-5pt} 
We proposed a novel FC-SCINet-based equalizer for downstream 100G PAM-4 PON with 28.7 dB OPL. FC-SCINet demonstrates remarkable tolerance to CD, achieving a minimum of 87.5\% BER improvement at 11 km transmission compared to a 9/21-tap FFE and a 3-layer DNN. The architecture of FC-SCINet excels in capturing and effectively addressing both spectral and temporal pulse distortions caused by CD. In the presence of EAM chirp, residual time jitter, and Kerr nonlinearity, FC-SCINet exhibits remarkable robustness, achieving an 88.87\% improvement in the BER at 5 km compared to FFE and DNN while offering 10.577\% lower complexity.
\vspace{-5pt} 

\section{Acknowledgement}
\vspace{-5pt}
Work funded by the Federal Ministry of Education and Research (BMBF) (KIGLIS: 16KIS1228, 16KIS1230).
\vspace{-5pt} 



\end{document}